# Effects of Integrated Heatsinks on Superconductivity in Tantalum Nitride Nanowires at 300 Millimeter Wafer Scale

Ekta Bhatia[1†*], Tharanga R. Nanayakkara[2†], Chenyu Zhou[2], Tuan Vo[2], Wenli Collison[1], Jakub Nalaskowski[1], Stephen Olson[1], Soumen Kar[1], Hunter Frost[3], John Mucci[1], Brian Martinick[1], Ilyssa Wells[1], Thomas Murray[1], Corbet Johnson[1], Charles T Black[2], Mingzhao Liu[2*], Satyavolu S Papa Rao[1*]

[1]NY Creates, Albany, NY 12203, USA

[2]Center for Functional Nanomaterials, Brookhaven National Laboratory, Upton, NY 11973

[3]College of Nanotechnology, Sci. & Eng., University at Albany (SUNY), Albany, NY 12203, USA

† These authors contributed equally to this work.

* Authors to whom correspondence should be addressed.

E-mail: ebhatia@ny-creates.org, spaparao@ny-creates.org, mzliu@bnl.gov

We report the superconducting properties of tantalum nitride (TaN) nanowires and TaN/copper (TaN/Cu) bilayer nanowires fabricated on 300 mm silicon wafers using CMOS-compatible processes. We evaluate how an integrated Cu heatsink modifies the superconducting response of TaN nanowires by improving thermal dissipation without significantly compromising key superconducting parameters. Through analysis of hysteresis in current-voltage curves, we demonstrate that Cu integration improves heat dissipation, supporting expectations of faster reset times in superconducting nanowire single-photon detectors (SNSPDs), consistent with enhanced heat transfer away from the hot spot. Using the Skocpol-Beasley-Tinkham (SBT) hotspot model, we quantify the Cu-enabled improvement in heat transfer as an approximately 100× increase in the SBT slope parameter β and effective interfacial heat-transfer efficiency compared to TaN nanowires. The near-unity ratio of critical to retrapping current in TaN/Cu bilayer nanowires provides another evidence of efficient heat removal enabled by the integrated Cu layer. Our results show a zero-temperature Ginzburg-Landau coherence length of 7 nm and a critical temperature of 4.1 K for 39 nm thick TaN nanowires. The nanowires show <5% variation in critical dimensions, room-temperature resistance, residual resistance ratio, critical temperature, and critical current across the 300 mm wafer for all measured linewidths, demonstrating excellent process uniformity and scalability. These results indicate the trade-offs between superconducting performance and heat-sinking efficiency in TaN/Cu bilayer nanowires. They also underscore the viability of wafer-scale fabrication for fast, large-area SNSPD arrays for applications in photonic quantum computing, cosmology, and neuromorphic computing devices.

## I. Introduction

Superconducting nanowire single-photon detectors (SNSPDs) have applications in quantum information science (including quantum communication) and astronomy [1-5], leveraging superconductivity to detect individual photons with high efficiency and timing resolution. Key performance metrics include high detection efficiency, low dark-count rate, broad spectral sensitivity, and short reset times. CMOS fab-compatible materials, processes and tools are likely to enable these metrics to be targeted by devices used in quantum computing or detector arrays that are readily scaled to high volumes in 300 mm wafer fabrication facilities.

The choice of material plays a crucial role in SNSPD performance. Niobium nitride (NbN) has traditionally been the material of choice due to its relatively high critical temperature ($T_c$) and ability to remain superconducting in ultrathin films (~3-4 nm) [6,7], which supports robust operation and simplifies fabrication. By contrast, TaN has a lower $T_c$ that reduces the energy barrier for forming photon-induced hotspots and enables easier suppression of superconductivity, increasing the probability of detecting lower-energy infrared photons and, hence, improving detection efficiency at longer wavelengths [2,9]. Additionally, its compatibility with CMOS-IC fabs [8] and demonstrated performance in ultrathin TaN films for SNSPDs make it an attractive alternative [9,10]. However, managing heat dissipation in superconducting nanowires, independent of the specific material system, remains a critical challenge for minimizing thermal retrapping and reducing reset times in SNSPDs. This motivates investigations into integrated metallic heatsinks, such as Cu, to improve heat dissipation without compromising superconductivity.

Previous reports have shown how sputtered TaN film properties depend on process parameters such as Ar/$N_2$ flow rates, sputtering power, and substrate temperature [8, 11-15]. TaN films also exhibit a superconductor-to-insulator transition as a function of film stoichiometry and thickness [16]. However, systematic investigations of TaN nanowires, particularly their superconducting characteristics in CMOS-compatible 300 mm wafer processes, are far from numerous [17]. Understanding these factors is critical for developing engineered bilayer designs, such as TaN/Cu, that optimize superconductivity and heat dissipation for faster reset quantum devices using scalable processes.

In this paper, we characterize and compare TaN/Cu bilayer and TaN nanowires fabricated at 300 mm wafer-scale to evaluate trade-offs between enhanced heat dissipation and changes in superconducting behavior. We determine $T_c$, critical current ($I_c$), and the zero-temperature Ginzburg–Landau coherence length ($\xi(0)$) for TaN and TaN/Cu bilayer nanowires across different nanowire widths, and quantify across-wafer uniformity of these parameters. We use the ratio of $I_c$ to retrapping current ($I_r$) and an effective interfacial heat-transfer efficiency extracted from fits to the Skocpol–Beasley–Tinkham (SBT) hotspot model as metrics to assess the effect of Cu. This comparison is enabled by two integration schemes: a chemical mechanical planarization (CMP)-based TaN/Cu bilayer process and a reactive ion etch (RIE)-based TaN-only process on Si/$SiO_2$.

The paper is organized as follows: Section II describes the two fabrication flows utilized, while Section III presents the room-temperature and cryogenic electrical characterization, and heat transport analyses using SBT hotspot modeling and $I_c$/$I_r$ to quantify the effect of Cu. Section IV summarizes the key results of our work.

## II. Experimental details

TaN thin films with a nominal thickness of 35 nm and stoichiometry TaN$_{0.53}$ (N/Ta = 0.53) were deposited on 300 mm Si (100) wafers by reactive sputtering from a Ta target in a 300 mm cluster tool. The actual TaN thickness is 39 nm, as determined from cross-sectional TEM for both TaN/Cu bilayer and TaN nanowires, and this value is used throughout this manuscript. To directly compare nanowires with and without Cu encapsulation, nanowires with widths ranging from 100 nm to 3 μm were fabricated using two integration schemes (Fig. 1). All widths quoted in the text are nominal (design) widths.

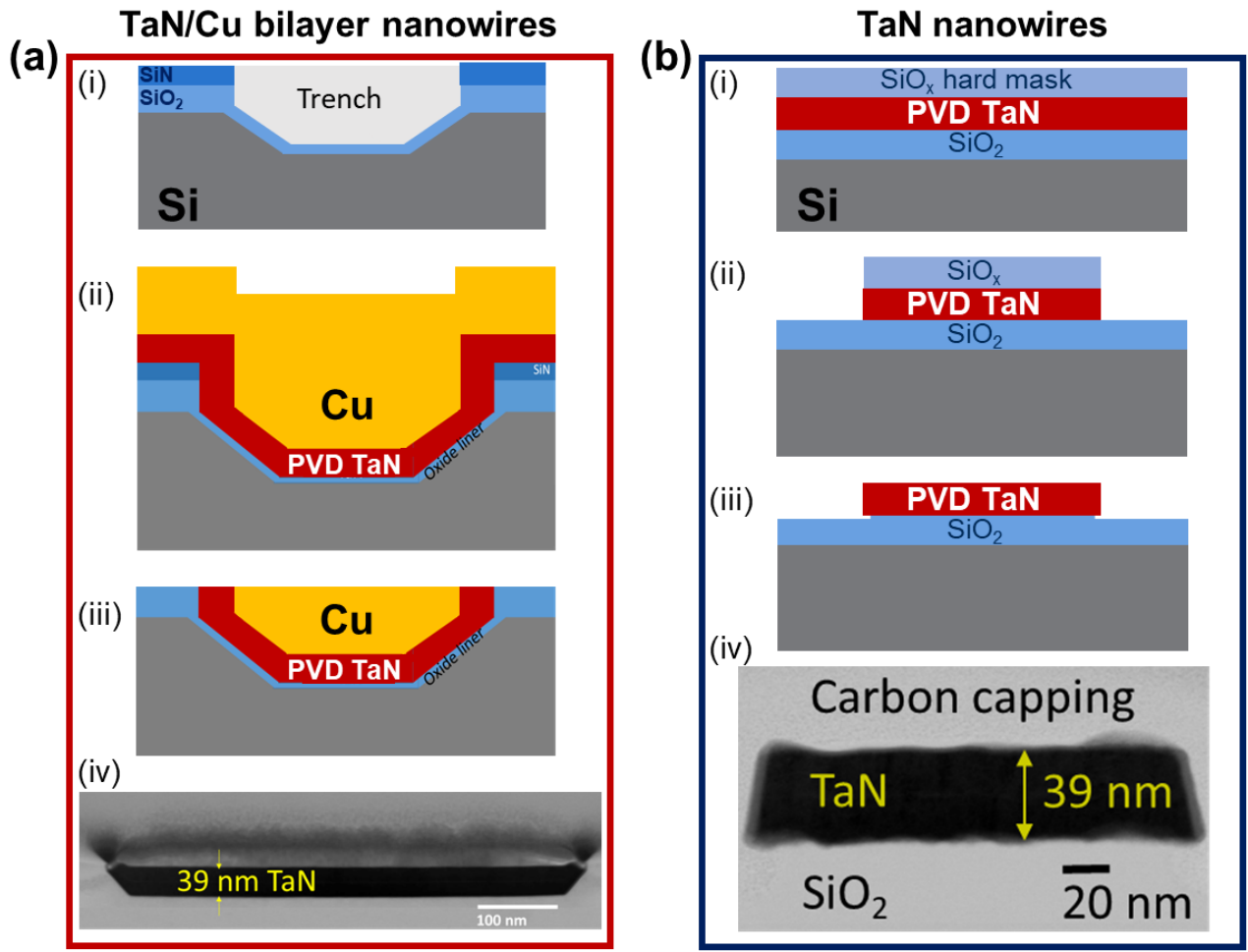

FIG. 1. Fabrication flows and cross-sectional TEM for (a) TaN/Cu bilayer nanowires and (b) TaN nanowires. The TaN/Cu process flow involves (i) trench patterning and oxide growth, (ii) in-situ PVD TaN/Cu seed deposition and Cu electroplating, and (iii) CMP. The TaN-only flow uses (i) blanket PVD TaN and sacrificial oxide deposition followed by (ii) 193 nm lithography, RIE, and (iii) wet etching. TEM confirms a TaN thickness of ≈39 nm in both structures. *(Red and blue borders denote TaN/Cu bilayer and TaN nanowire samples, respectively. This color scheme is used consistently throughout this work.)*

The TaN/Cu bilayer nanowires were fabricated using trench patterning followed by oxide growth in the trench and deposition of PVD TaN and in-situ Cu seed, Cu electroplating, and CMP. The TaN-only nanowires were fabricated on 50 nm thermally grown SiO₂, followed by blanket PVD TaN and SiOx (sacrificial oxide that acts as a hard mask for TaN during RIE) deposition, 193 nm lithography, RIE, and wet etching to remove sacrificial oxide. Detailed process flows and CMP design rules are provided in Refs. [17-20]. The inclusion of Cu necessitates the use of CMP, as no viable RIE process for Cu exists.

The characterization of blanket TaN films including X-ray diffraction (XRD), transmission electron microscopy (TEM), secondary ion mass spectrometry (SIMS), and X-ray photoelectron spectroscopy (XPS) for N/Ta ratios spanning 0.35–0.70 was reported in our prior work [17]. The room temperature resistances were measured at 26 dies (12.5 × 15.5 mm² each) across the wafer, with two nominally identical copies of each width per die, critical dimension (CD) measurements being made at one location per die. Additional cryogenic characterization of patterned TaN/Cu bilayer nanowires for these different N/Ta ratios is provided in the Supplementary Information. The film with stoichiometry N/Ta = 0.53 is the primary composition analyzed for both room-temperature and cryogenic measurements in the present study.

## III. Results and Discussion

The CD of features, designed as 110 nm wide lines or trenches, is highly uniform across the 300 mm wafer at each fabrication step, with within-wafer non-uniformity (WIWNU = σ/median × 100) <1% for both TaN/Cu bilayer and TaN CD lines (Fig. 2).

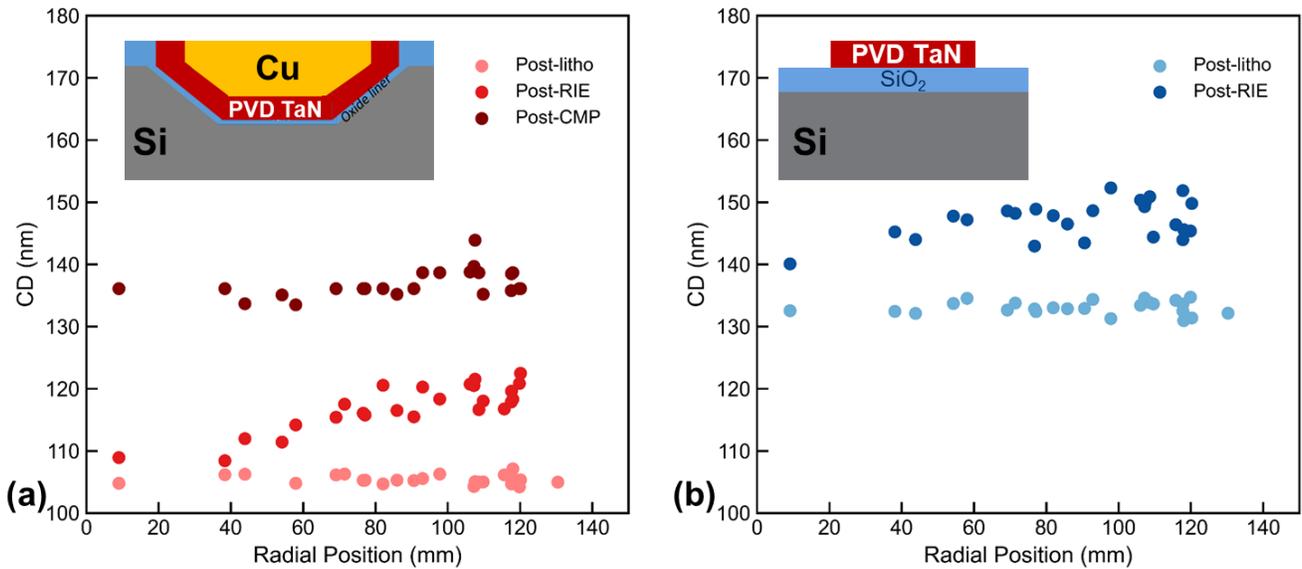

FIG. 2. Radial dependence of nanowires with critical dimension (CD) of 110 nm across the 300 mm wafer for (a) TaN/Cu bilayer and (b) TaN nanowires at various process steps. TaN/Cu: post-lithography, post-RIE, post-CMP; TaN: post-lithography, post-RIE. Insets show schematic cross-sections of the corresponding stacks.

Room-temperature resistance decreases with increasing nanowire width for both TaN/Cu bilayer and TaN nanowires (Fig. 3(a,f)). Across-wafer resistance is highly uniform for TaN (WIWNU < 4% for all widths). TaN/Cu bilayer nanowires show broader distributions (WIWNU decreasing from 14% to 10% as width increases) (Fig. 3(b,g)). Although CD uniformity is better than 1%, the resistance WIWNU is slightly higher because resistance samples thickness (d) variations [17] in addition to width. The larger WIWNU for TaN/Cu is attributable to CMP-induced Cu-thickness non-uniformity across the wafer, which impacts only the Cu layer. However, Cu-thickness-driven resistance variation is irrelevant below the TaN superconducting transition temperature (since it is shunted by the superconducting TaN). Figures 3(c-e, h-j) show the radial dependence of resistance for the narrowest and widest nanowires, along with representative top-view SEM images.

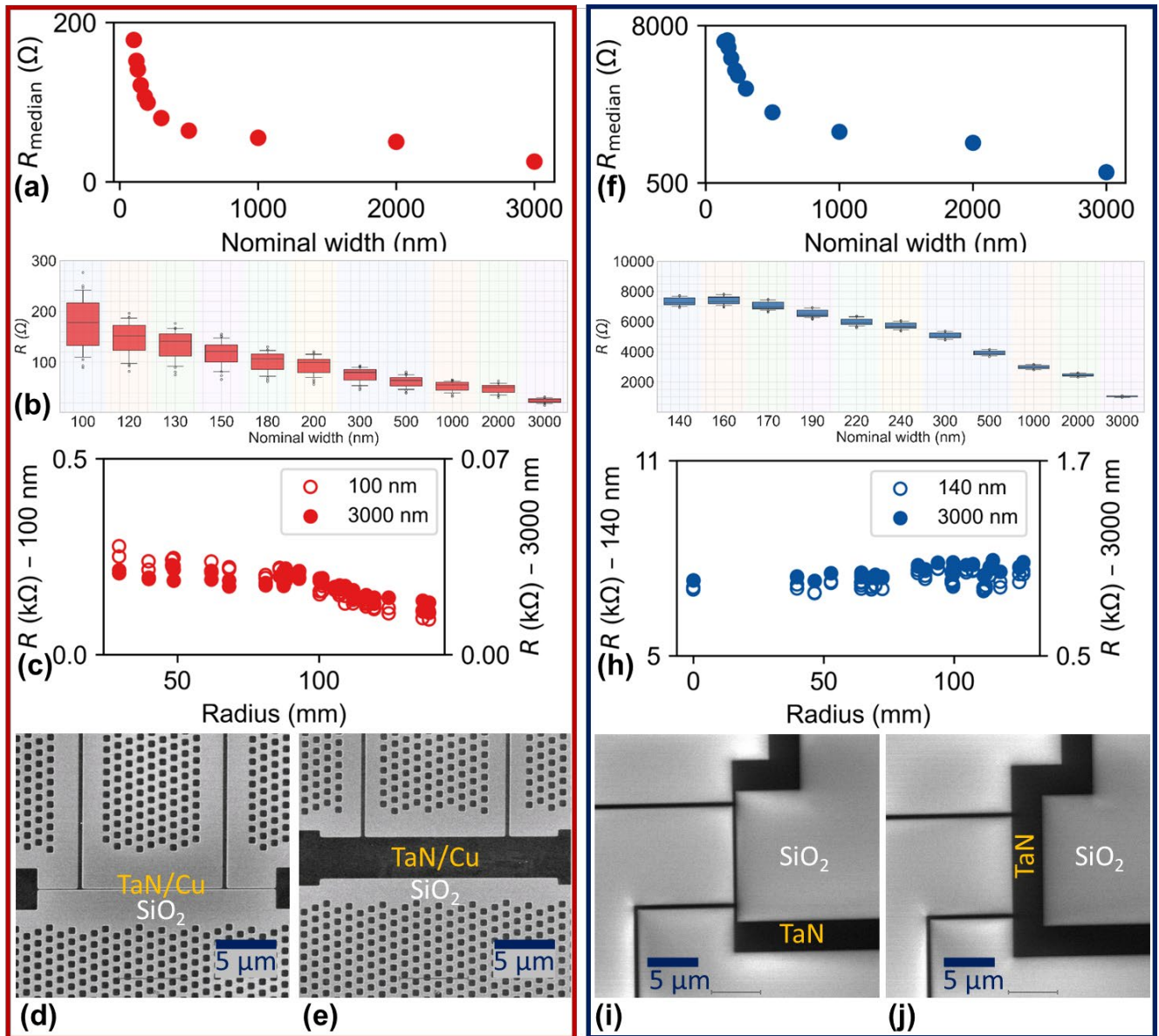

FIG. 3. Room-temperature electrical characterization and top-view SEM images of TaN/Cu bilayer nanowires (left column, (a–e)) and TaN nanowires (right column, (f–j)). (a,f) Median resistance vs nominal nanowire width. (b,g) Resistance distributions by width (box-and-whisker: boxes = interquartile range; whiskers = 5th–95th percentiles; circles outside of whiskers = outliers). Data are from 52 nanowires per width measured across 26 dies uniformly distributed over the 300-mm wafer. (c,h) Radial dependence of resistance for the narrowest and widest widths. (d,e,i,j) Representative top-view SEM images of the narrowest and widest nanowires and leads; darker contrast indicates TaN (or TaN/Cu) and lighter contrast indicates $SiO_2$.

TaN serpent lines show nearly geometry-independent sheet resistance, with 77-83 Ω/sq across all line widths (LW) and line spacings (LS) with < 8% variation (Fig. 4). Figure 4(a) shows an SEM image of the serpent line between comb electrodes used for the $R_s$ measurement. Figure 4(b) plots $R_s$ as a function of line width (LW) for line spacings (LS) of 300 nm and 1 μm. At fixed LW, $R_s$ is slightly lower for LS = 1 μm than for LS = 300 nm. This modest pattern-density dependence (higher $R_s$ for denser patterns) is attributed to RIE microloading, a well-known phenomenon in plasma etching, where local open area controls the available reactive flux and affects the etched profile [22]. This results in a reduced effective conducting cross-section, leading to higher sheet resistance $R_s$ for denser patterns. The room-temperature resistivity of TaN nanowires is 327 μΩ·cm, calculated from the $R_s$ value for the 3 μm linewidth shown in Fig. 4(b).

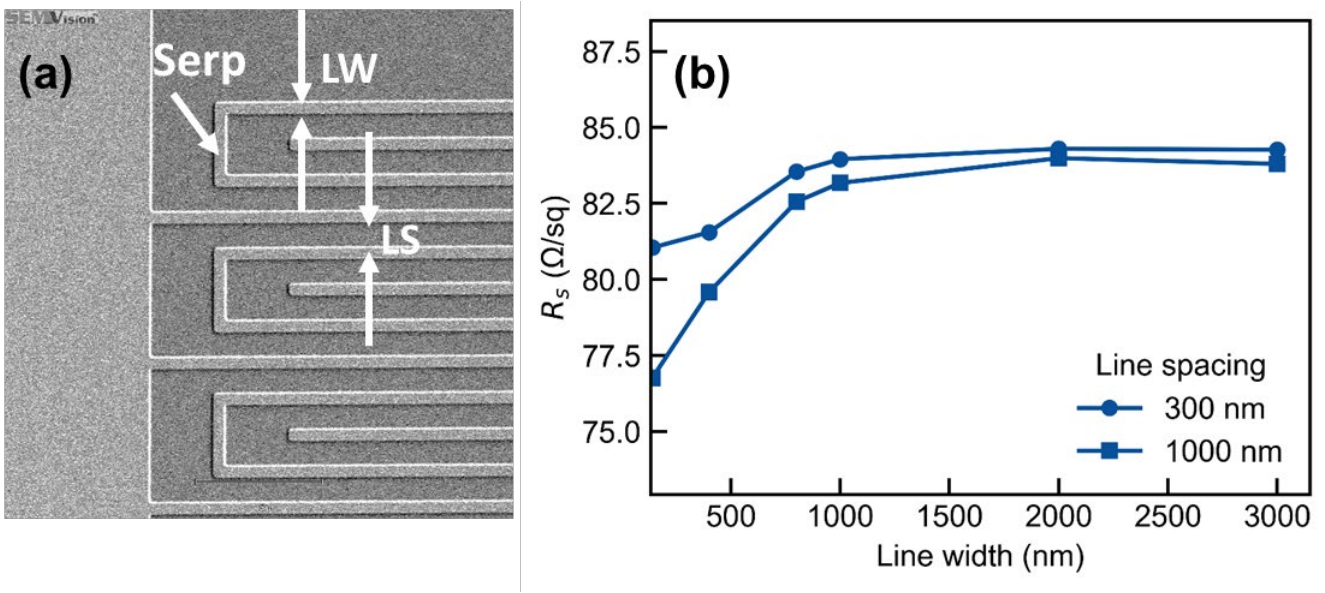

FIG. 4. Sheet resistance of TaN. (a) SEM image of the serpent between comb lines. LW refers to line width of TaN in the serpent and LS refers to the line spacing between two TaN lines (one from the serpent and one from the comb). (b) Sheet resistance as a function of line width for line spacings of 300 and 1000 nm.

After confirming across-wafer repeatability at room temperature, we characterized selected chips at cryogenic temperatures. Cryogenic $R(T)$ measurements show distinct normal state behavior for the two structures and a reduced $T_c$ in TaN/Cu bilayer nanowires (Fig. 5). $T_c$ is defined as the midpoint transition temperature $T_c^{mid}$, obtained from the peak in $dR/dT$ (see Supplementary S3). The resistance is dominated by the high conductivity of Cu, resulting in a decrease in resistance on cooling in TaN/Cu bilayer nanowires. Conversely, TaN nanowires deposited on $SiO_2$ show an increase in resistance upon cooling, consistent with existing literature on PVD TaN [16]. The residual resistance ratio (RRR), defined as $R_{300K}/R_{6K}$ (with 6 K in the normal state, just above $T_c$), is ~0.9 for TaN nanowires and 1.56 for TaN/Cu bilayer nanowires (dominated by Cu). The inset of Fig. 5 highlights the superconducting transition and shows that Cu encapsulation reduces $T_c$ (due to the superconducting proximity effect at the TaN/Cu interface).

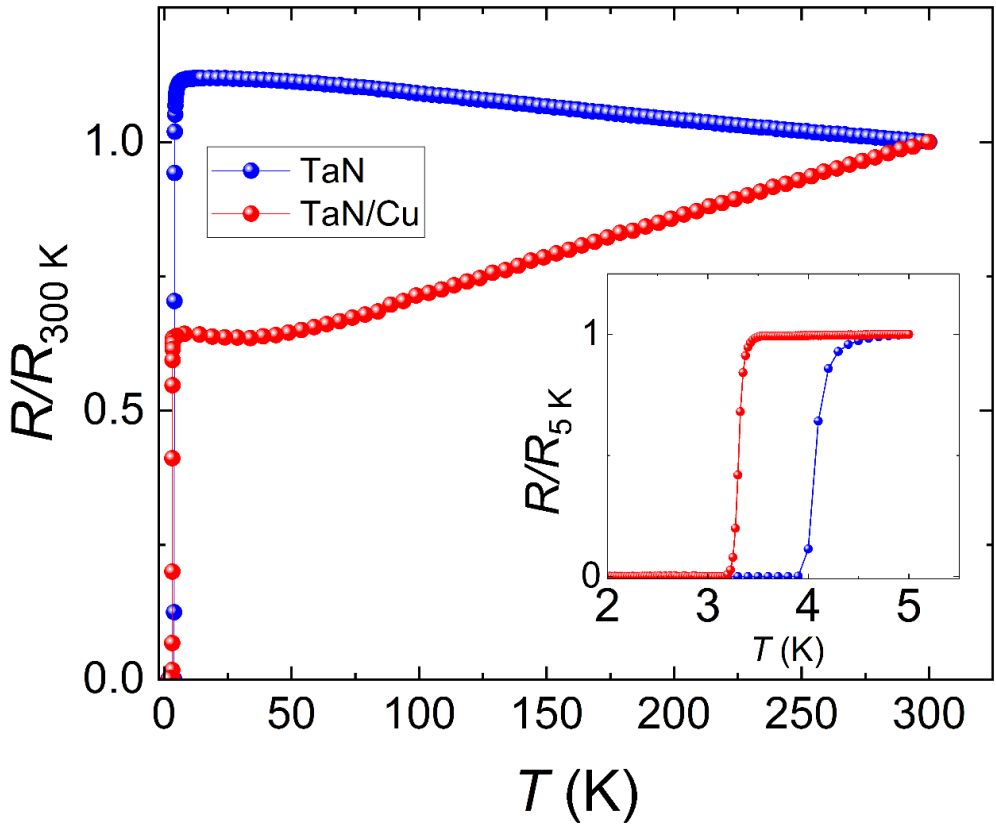

FIG. 5. Normalized resistance (resistance normalized to $R_{300\,K}$) versus temperature for TaN/Cu bilayer nanowires (red) and TaN nanowires (blue). The inset shows a zoomed-in view near the superconducting transition (resistance normalized to $R_{5\,K}$)

Both TaN/Cu bilayer and TaN nanowires show $T_c$ that is nearly independent of nanowire width (Fig. 6). The transition width, defined as the temperature interval between 10% and 90% of the normal-state resistance, is broader for narrower nanowires due to their higher surface-to-volume ratio and increased sensitivity to edge effects and disorder. This broadening is more pronounced in TaN/Cu bilayer nanowires, likely arising from the superconducting proximity effect.

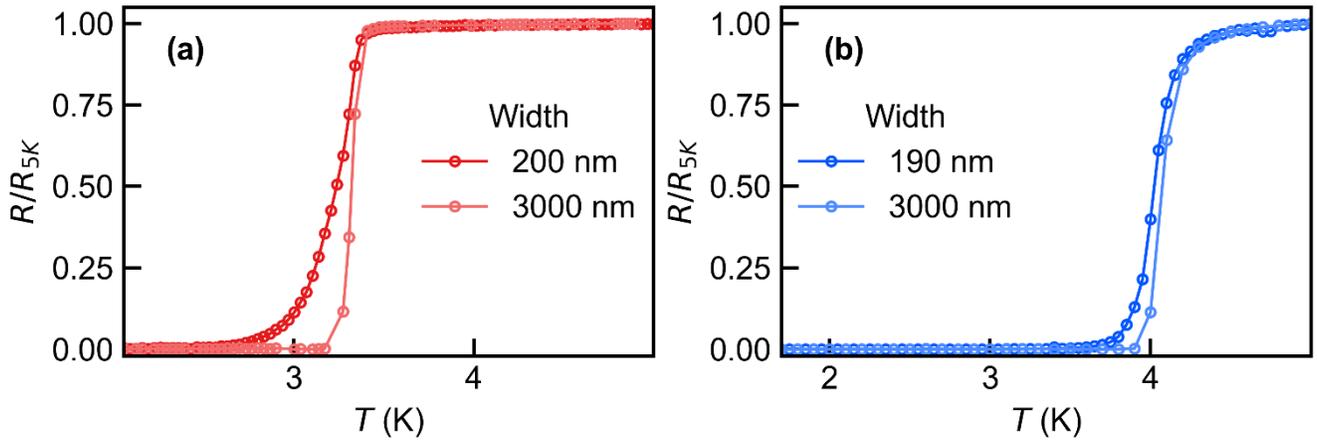

FIG. 6. Normalized resistance as a function of temperature for (a) TaN/Cu bilayer nanowires with nominal widths of 200 nm and 3 μm and (b) TaN nanowires with nominal widths of 190 nm and 3 μm. In each case, resistance is normalized to its value at 5 K.

For all nanowire widths, $T_c$ is highly uniform across the 300 mm wafer for TaN/Cu bilayer and TaN nanowires (Fig. 7). The across-wafer variation is <5%, demonstrating excellent process uniformity needed for scalable fabrication. Consistent with Fig. 5, TaN/Cu bilayers exhibit a systematically reduced $T_c$ relative to TaN-only nanowires, but without loss of across-wafer uniformity.

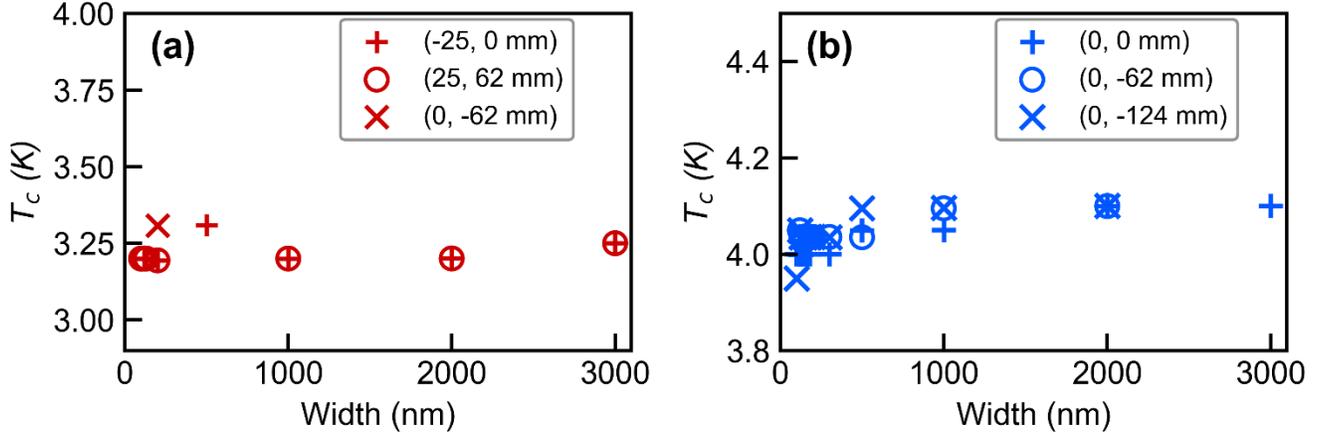

FIG. 7. Across wafer uniformity of $T_c$ for (a) TaN/Cu bilayer nanowires and (b) TaN nanowires. The different symbols denote different wafer locations (coordinates in mm; wafer center at (0, 0)).

RRR and kinetic inductance per square ($L_{K,\square}$) are uniform across the 300 mm wafer for TaN nanowires (Fig. 8). As shown in Figs. 8(a) and 8(b), both RRR and $L_{K,\square}$ remain nearly constant across nanowire widths and wafer positions, indicating good across-wafer uniformity. The zero-temperature kinetic inductance per square, $L_{K,\square}(0)$, is related to the normal-state $R_s$ (measured above $T_c$) by [32]:

$$L_{K,\square}(0) = \hbar R_s / \pi \Delta_0,$$

where $\Delta_0 = 1.76 k_B T_c$. Using the measured $R_s$ values at 300 K and RRR to estimate $R_s$ at 6 K, we calculate $L_{k,\square}(0) = 31.62 \pm 0.5 \mathrm{pH}/\square$. Such kinetic inductance values can be advantageous for traveling-wave parametric amplifiers (TWPAs) and compact lumped-element resonators, as they enable device miniaturization. For wafer locations away from the center, $R_s$ was taken from the nearest available serpent structure measurement at room temperature.

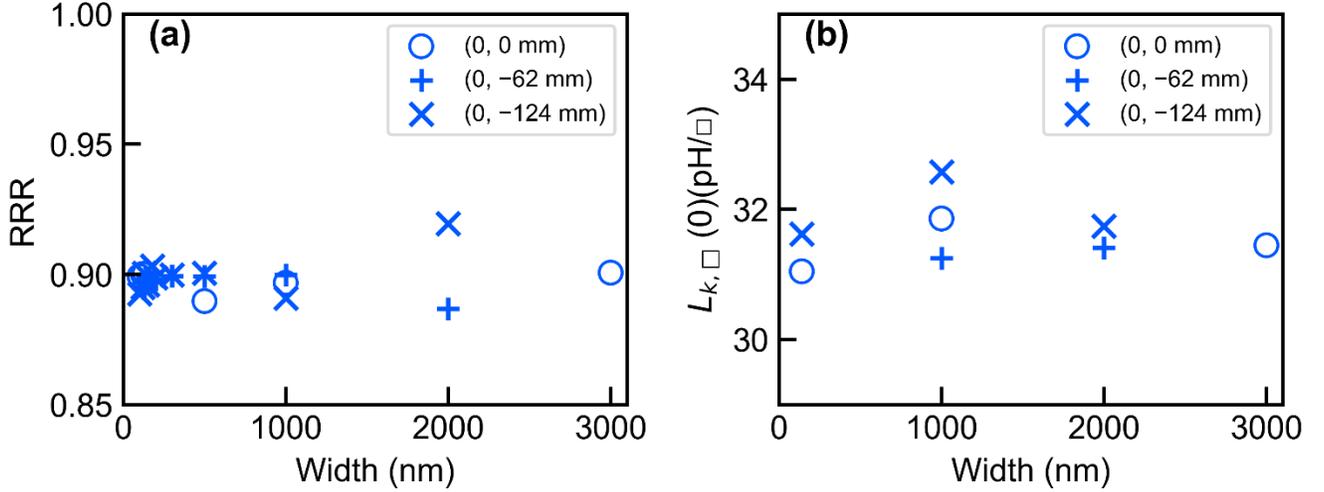

FIG. 8. Across-wafer dependence of the RRR and zero-temperature kinetic inductance per square ($L_{K,\square}(0)$ for TaN nanowires. (a) RRR vs width (b) $L_{K,\square}(0)$ vs width. Symbols denote different wafer locations (coordinates in mm).

Figure 9 is used to extract the upper critical field $H_{c2}(T)$ and determine $\xi(0)$ as a function of nanowire width for TaN/Cu bilayer (left panels) and TaN (right panels) nanowires. Representative $R(H)$ curves at fixed temperatures for 3 μm wide nanowires are shown in Figs. 9(a) and 9(d), and the corresponding normalized resistance maps $R/R_n(T,H)$ are shown in Figs. 9(b) and 9(e), where $R_n$ is the normal state resistance just above the transition. In both structures, the resistance is near zero at low field and increases monotonically with magnetic field through the superconducting to normal transition. The transition shifts to lower field as temperature increases, consistent with temperature-dependent suppression of superconductivity. In TaN/Cu bilayer nanowires, the high field normal state resistance shows a weak negative magnetoresistance due to Cu [21]. Additional $R/R_n(T,H)$ maps for narrower nanowires are shown in Figs. 9(c) and 9(f). We define $H_{c2}(T)$ using the $R/R_n = 0.5$ contour (white boundary) and extract $dH_{c2}/dT|_{T_c}$ from a linear fit near $T_c$ (dashed line). $H_{c2}(0)$ is estimated using the Werthamer–Helfand–Hohenberg (WHH) relation [23,28,29]:

$$H_{c2}(0) = -\frac{\pi T_c}{8e^\gamma}\frac{dH_{c2}}{dT}\Big|_{T=T_c} \quad (1.1)$$

Where γ is Euler–Mascheroni constant. The Ginzburg–Landau coherence length, ξ(0) is then calculated from

$$\mu_0 H_{c2}(0) = \frac{\Phi_0}{2\pi\xi^2(0)} \quad (1.2)$$

where $\mu_0$ is the vacuum permeability and $\Phi_0$ is the magnetic flux quantum. The resulting coherence lengths for 3 μm wide nanowires are 8.4 nm for TaN/Cu bilayer and 7 nm for TaN. These values are comparable to the 5 nm coherence length previously reported for high temperature deposited TaN films on sapphire [24], and hence demonstrate that superconducting TaN films deposited on Si substrates can be adopted by high volume fabrication facilities.

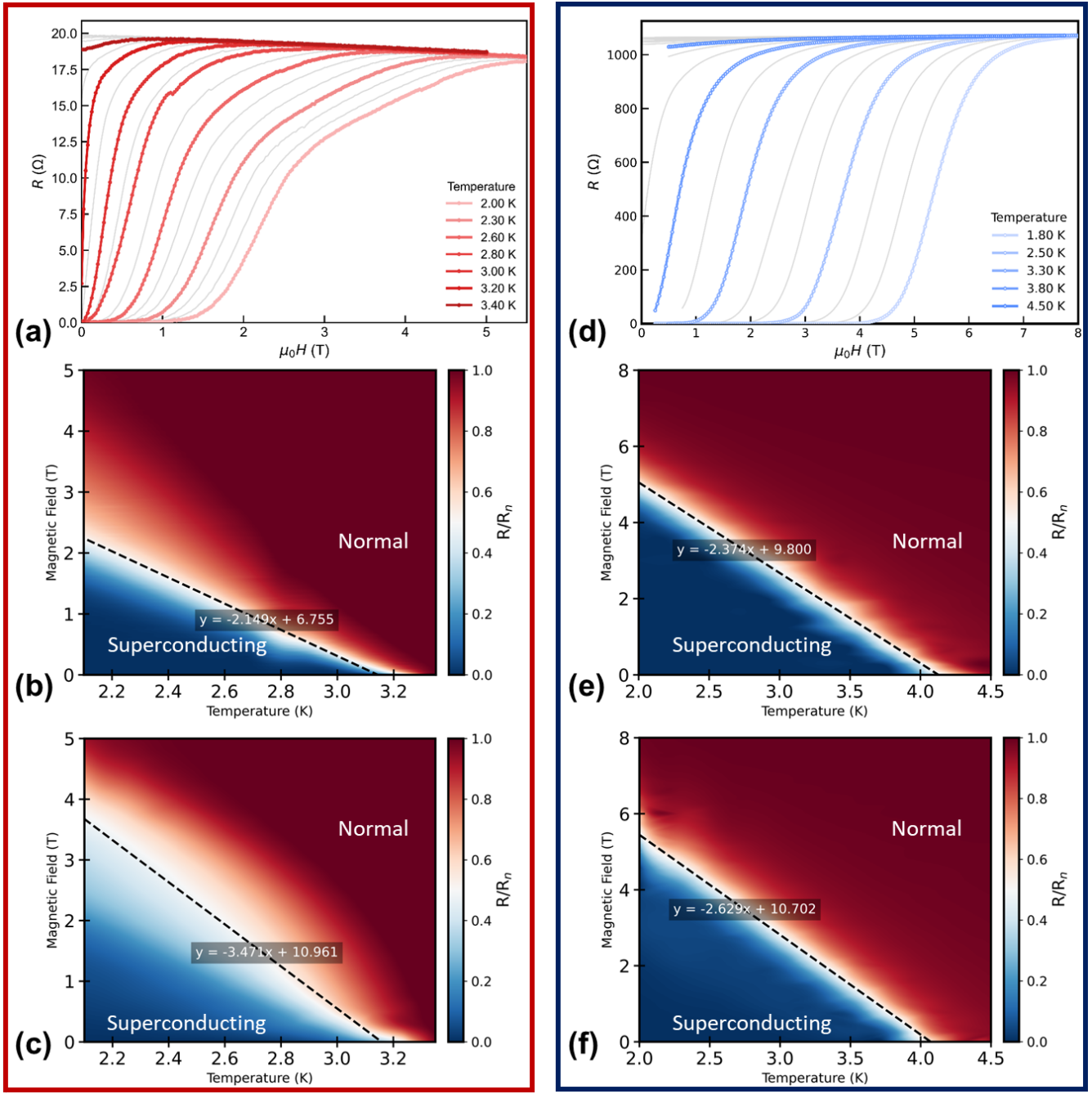

FIG. 9. Extraction of the upper critical field $H_{c2}(T)$ and coherence length $\xi(0)$ from magnetic field dependent superconducting transitions for TaN/Cu bilayer nanowires (a–c) and TaN nanowires (d–f). (a,d) $R(H)$ at several temperatures for 3 μm wide nanowires. (b,e) Corresponding 2D maps of normalized resistance $R/R_n(T, H)$, where $R_n$ is the normal-state resistance just above the superconducting transition. (c) 2D map for a 200 nm wide TaN/Cu bilayer nanowire. (f) 2D map for a 140 nm wide TaN nanowire. The white boundary denotes $R/R_n = 0.5$ contour used to define $H_{c2}(T)$; dashed lines are linear fits near $T_c$ used to extract $dH_{c2}/dT |_{T_c}$ for estimating $H_{c2}(0)$ and $\xi(0)$.

Panels (c) and (f) show the 2D maps for 200 nm wide TaN/Cu bilayer and 140 nm wide TaN nanowires, respectively. For TaN nanowires, the slope $dH_{c2}/dT |_{T_c}$ and the extracted coherence length across widths are similar (within 5%), consistent with the uniform $T_c$ observed across the wafer. In contrast, for TaN/Cu bilayer nanowires, the magnitude of $dH_{c2}/dT |_{T_c}$ increases, yielding a higher estimated $H_{c2}(0)$ and a shorter coherence length of ~6.6 nm. This reduction may arise from two effects: (i) the broader transition in TaN/Cu, where a wider $R/R_n$ boundary (e.g., panel (c)) increases uncertainty in extracting $dH_{c2}/dT |_{T_c}$ from the $R/R_n = 0.5$ contour, and (ii) stronger proximity-induced pair breaking in narrower bilayers due to the increased influence of the TaN/Cu interface [17].

Figure 10 shows the I–V characteristics of TaN/Cu bilayer and TaN nanowires of different widths measured at zero magnetic field. TaN/Cu bilayer nanowires (Fig. 10(a)) exhibit negligible hysteresis, with $I_c \approx I_r$, consistent with Cu extracting heat from the nanowire region and acting as an effective on-chip heat-sink [25,26]. In contrast, TaN nanowires (Fig. 10(b)) show pronounced hysteresis, due to weaker thermal coupling to the substrate. We note that the critical current ($I_c$) is defined as the switching current from the superconducting to the normal state on the up-sweep, while the retrapping current ($I_r$) is the return transition to the superconducting state on the down-sweep.

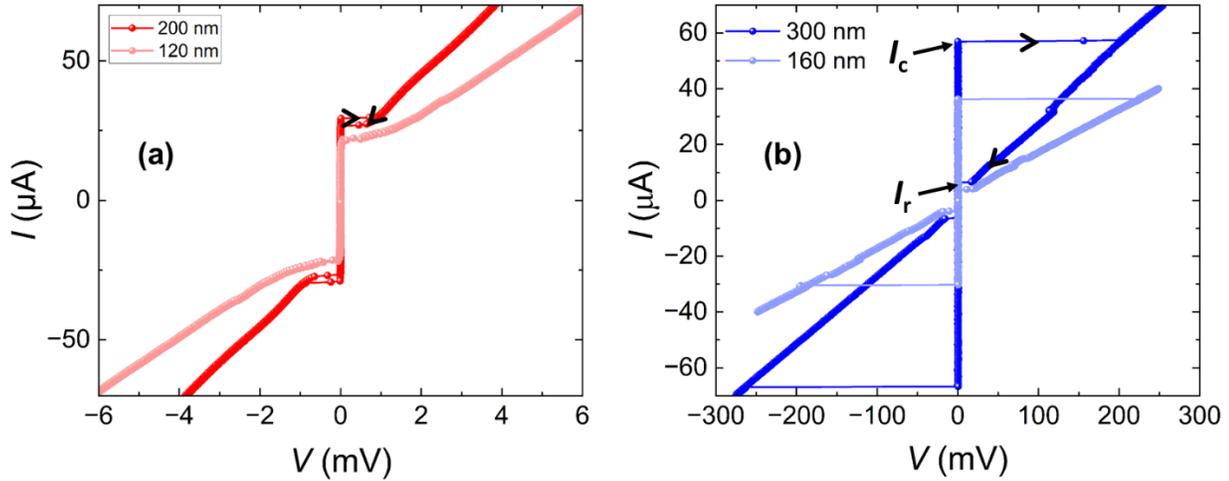

FIG. 10. I–V characteristics at zero magnetic field for (a) TaN/Cu bilayer nanowires and (b) TaN nanowires (linewidths indicated in the legends; note the different voltage scales). The critical current $I_c$ and retrapping current $I_r$ are indicated by black arrows.

Figure 11 shows $I_c$ as a function of nanowire width for TaN/Cu bilayer and TaN nanowires measured at multiple wafer locations. In both structures, $I_c$ increases with width, and the across-wafer variation among the measured sites is <5%.

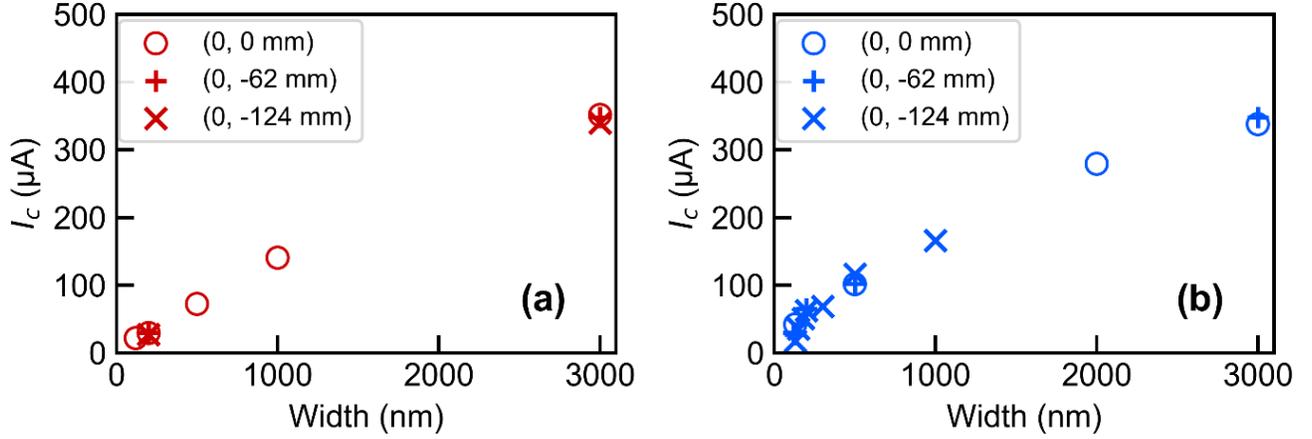

FIG. 11. Critical current ($I_c$) versus nanowire width for (a) TaN/Cu bilayer nanowires and (b) TaN nanowires measured at multiple wafer locations (symbols denote coordinates in mm; wafer center at (0, 0)). In both structures, $I_c$ scales with width and shows <5% variation across the measured wafer sites.

Copper enhances hotspot cooling in TaN nanowires. This is evidenced by the retrapping current analysis of 3 μm wide devices, where the TaN/Cu bilayer exhibits an approximately 100× larger SBT slope parameter β than bare TaN (Fig. 12). We extract the temperature dependent hotspot current $I_{hs}(T)$ from the constant-current plateau on the down-sweep of the I–V characteristics at finite voltage (Fig. 12(a)), following Dane et al. [26]. In our devices, this plateau current is typically equal (within measurement uncertainty) to the retrapping current $I_r$ defined as the current at which the device returns to the superconducting branch. We restrict the analysis to temperatures where the I–V curves are hysteretic and the plateau is unambiguous. In this regime, the Skocpol–Beasley–Tinkham (SBT) hotspot model can be applied, and $I_{hs}$ (T) follows [25,26]:

$$I_{hs}(T) = A\sqrt{T_{hs}^n - T^n}$$

where $A$ is a geometry dependent prefactor, $T_{hs}$ is the hotspot temperature at which $I_{hs}$ becomes zero, indicating disappearance of a stable hotspot state. and $n$ captures the dominant heat-loss mechanism. For both TaN and TaN/Cu datasets, $n$ was fixed at 4, consistent with phonon-limited (Kapitza-like) cooling pathways (similar to that reported for NbN on different dielectric substrates including $SiO_2$) [26,30]. The fitted parameters $A$ and $T_{hs}$ are summarized in Table I.

To compare heat-removal efficiency more directly, we plot the linearized SBT form (Figs. 12(b,c)):

$$\frac{I_{hs}^2 R_s}{T_{hs}^4 w^2} \text{ versus } 1 - (T/T_{hs})^4,$$

where $R_s$ is the TaN sheet resistance in the normal state and $w$ is the nanowire width. Because the hotspot forms in the TaN layer for both structures, and because the measured normal resistance of the TaN/Cu bilayer is strongly shunted by Cu, we use the same TaN $R_s$ value for both datasets to avoid distortions from Cu shunting. The resulting linear fits yield the slope $\beta$, which is proportional to the effective interfacial heat-transfer strength [26,31] The ~100× larger $\beta$ for TaN/Cu provides quantitative evidence that Cu improves hotspot cooling.

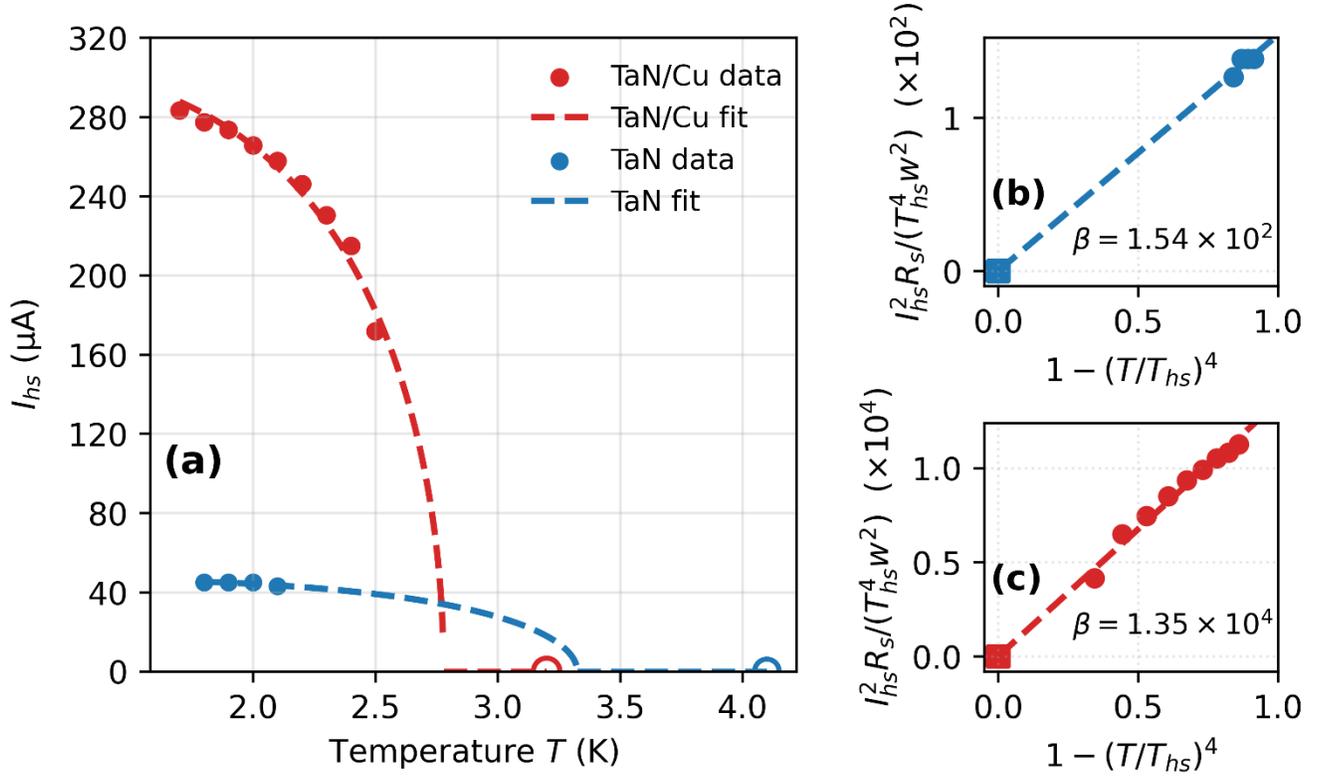

FIG. 12. Hotspot current analysis for 3 µm wide TaN and TaN/Cu nanowires. (a) $I_{hs}(T)$ extracted from hysteretic I-V curves (symbols) acquired at different temperatures with SBT hotspot model fits (dashed lines); open circles represent $T_c$. (b,c) The corresponding linearized SBT plots for TaN and TaN/Cu, yielding slope $\beta$ (effective interfacial heat-removal efficiency); the much larger $\beta$ for TaN/Cu indicates substantially enhanced hotspot cooling with the Cu overlayer. The origin marked by a square corresponds to $T = T_{hs}$, where $T_{hs}$ is extracted from the fit in (a), and listed in Table I.

**Table I**: Structural and superconducting parameters of 3 µm wide TaN/Cu bilayer and TaN nanowires. The N/Ta ratio was determined from sputter XPS analysis [19], and the film thickness ($d$) from TEM cross-sections (Fig. 1). The RRR and $T_c$ were determined from $R(T)$ measurements (Fig. 5). The critical current density ($J_c$) was calculated from the $I_c$ extracted from I–V curves and the nanowire cross-sectional area determined from TEM images using ImageJ. The Ginzburg–Landau coherence length $\xi_{GL}(0)$ was derived from the slope $|dH_{c2}/dT|_{T_c}$ obtained from R(H) traces using the WHH analysis described above. The lower $T_{hs}$ in TaN/Cu is consistent with Fig. 12 and may be related to the reduced $T_c$ due to proximity effect at the TaN/Cu interface [26].

| Nanowire type | N/Ta | d (nm) | RRR | $T_c$ (K) | $J_c$ (MA/cm²) | $\left|\frac{dH_{c2}}{dT}\right|_{T=T_c}$ (T/K) | $\xi_{GL}(0)$ (nm) | $R_s$ (Ω/sq) | A (AK⁻²) | $T_{hs}$ (K) |
|---|---|---|---|---|---|---|---|---|---|---|
| TaN/Cu | 0.53 | 39 | Cu-shunted | 3.2 | 0.39 | 2.2 | 8.4 | Cu-shunted | 4.0×10⁻⁵ | 2.8 |
| TaN | | | 0.9 | 4.1 | 0.52 | 2.4 | 7 | 84 | 4.3×10⁻⁶ | 3.3 |

Figure 13 shows the effect of Cu on heat transport efficiency by comparing the ratio $I_c/I_r$ of TaN/Cu bilayer and TaN nanowires. In the SBT hot-spot picture, hysteresis in the nanowire I–V curve arises from self-heating: when the bias current reaches $I_c$, the nanowire develops a resistive hot spot. This hot spot can persist as the current is reduced, so the device retraps only when the current falls below $I_r$. Accordingly, $I_c/I_r$ provides a useful proxy for heat-removal efficiency: large $I_c/I_r$ indicates strong self-heating (poor heat sinking), whereas $I_c/I_r \approx 1$ reflects efficient heat removal and negligible hysteresis. Across the measured width range at zero field, Cu encapsulation reduces $I_c/I_r$ from ~10-15 (TaN) to ~1-1.3 (TaN/Cu), demonstrating that the Cu overlayer functions as an effective on-chip heat sink.

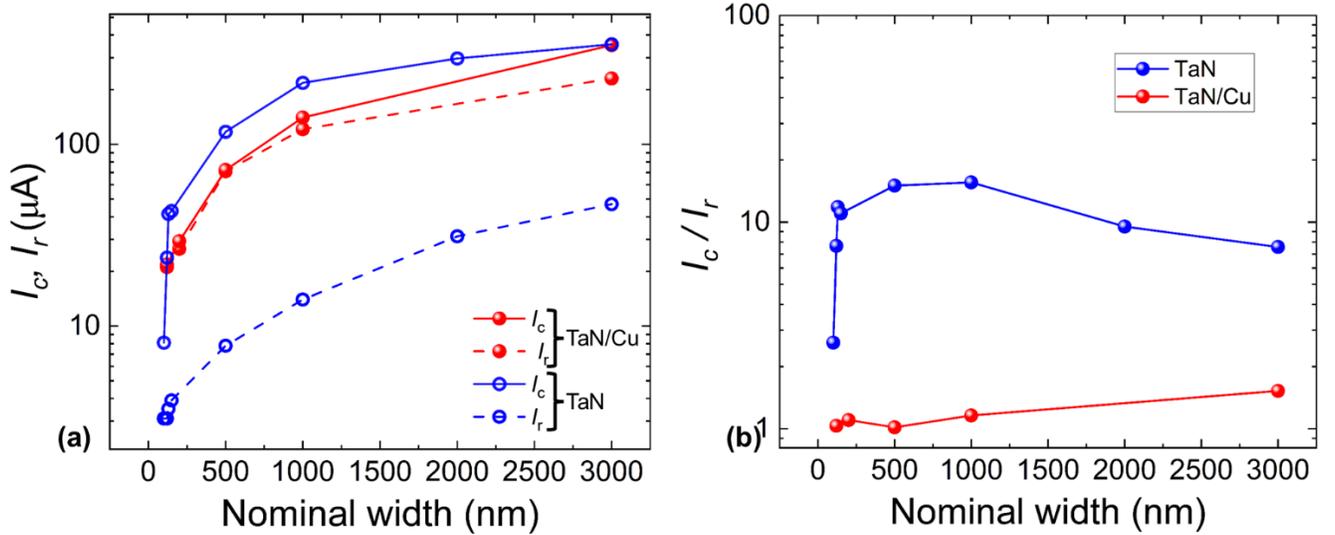

FIG. 13. (a) $I_c$ (solid) and $I_r$ (dashed) versus nominal nanowire width for TaN/Cu bilayer (red) and TaN (blue) nanowires, measured at zero magnetic field (T = 2 K). (b) Hysteresis metric $I_c/I_r$ versus width. The y-axis is plotted on a logarithmic scale.

Reducing the TaN thickness suppresses $T_c$ in both structures, with a stronger suppression in TaN/Cu due to the superconducting proximity effect (Fig. 14). All results presented so far were for 39 nm TaN nanowires. A lower $T_c$ can be advantageous for SNSPDs by lowering the hotspot formation threshold and extending sensitivity toward longer wavelengths. At the same time, Cu encapsulation improves thermal sinking and mitigates latching, consistent with the reduced hysteresis and larger SBT slope in Figs. 12–13. However, there is a caveat: if heat removal is excessively rapid and strong, the resistive perturbation from an absorbed photon may relax too quickly, potentially reducing detection efficiency [27]. Therefore, optimizing the Cu cross-section is important to balance fast reset with robust hot-spot formation. These trade-offs are evident in our data: Figs. 5 and 7 show the reduction of $T_c$ in Cu-encapsulated nanowires due to proximity effects, whereas Figs. 12 and 13 demonstrate the corresponding improvement in heat transport efficiency. In our process, the post-CMP Cu thickness depends on TaN thickness because the trench depth is fixed; optimizing the TaN/Cu geometries (including Cu thickness) will be explored in future work.

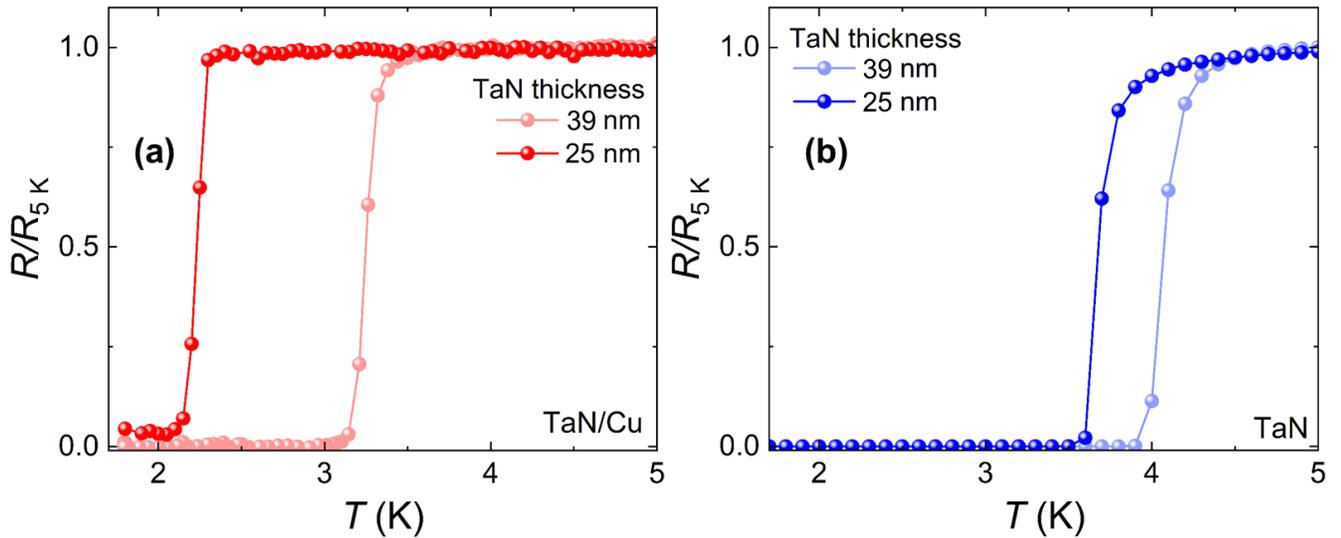

FIG. 14. Thickness dependence of the superconducting transition temperature for 3 μm wide nanowire at zero magnetic field: normalized $R(T)$ for (a) TaN/Cu bilayer nanowires and (b) TaN nanowires. $R$ is normalized to its value at 5 K.

## IV. Conclusions

This work establishes a fully CMOS-compatible, 300 mm wafer-scale process for fabricating ultrathin TaN nanowires and TaN/Cu bilayer nanowires with excellent dimensional, electrical, and cryogenic performance uniformity. The $T_c$ of TaN nanowires was 4.1 K, and the coherence length was 7 nm for TaN films deposited at room temperature by reactive magnetron sputtering. By systematically varying nanowire dimensions, thickness, and stoichiometry, and by integrating Cu, we demonstrated tunable $T_c$, $I_c$, and heat transport efficiency. The ratio $I_c/I_r$ served as a semi-quantitative metric for heat transport efficiency and decreased from 15 to 1 upon Cu integration. Thermal transport analysis using SBT hotspot modeling provides quantitative insight into how Cu overlayers modify hotspot cooling pathways in these disordered superconducting nanowires. These results show that Cu integration improved the extracted heat removal metric by approximately two orders of magnitude, providing an effective heat sink that could reduce reset times in SNSPDs.

Across-wafer non-uniformity in critical dimensions after each process step, electrical resistance, and cryogenic properties such as $T_c$ and $I_c$ was less than 5% for nanowire widths from 100 nm to 3 μm. The demonstrated process control, combined with tunability of superconducting and thermal properties, makes TaN and TaN/Cu nanowires promising candidates for next-generation quantum photonic and sensing technologies.

## V. Acknowledgments

This study was supported by the U.S. Department of Energy, Office of Science, National Quantum Information Science Research Centers, Co-design Center for Quantum Advantage (C2QA), under Contract No. DE-SC0012704, including Subcontract No. 390040. This research used the Physical Properties measurement system (PPMS) at the Center for Functional Nanomaterials (CFN), U.S. Department of Energy Office of Science Facilities at Brookhaven National Laboratory, under Contract No. DE-SC0012704.

# Supplementary Material for Effects of Integrated Heatsinks on Superconductivity in Tantalum Nitride Nanowires at 300 Millimeter Wafer Scale

## S1. Additional Fabrication and Metrology Details

Representative CD-SEM images confirm distinct etch/CMP profiles for TaN/Cu (post-CMP) and TaN (post-wet-etch) test lines at three wafer locations (Fig. S1). These CD lines were patterned using the same lithography and etch/CMP steps as the nanowires and therefore reflect the corresponding process profiles. The bright edge contrast in the TaN CD lines arises from the trapezoidal cross-section produced by reactive ion etching (RIE): the sloped sidewall creates a bright shoulder at the wire edges in top-view imaging. After CMP, the TaN/Cu bilayer CD lines exhibit smoother surfaces with visible Cu grain structure.

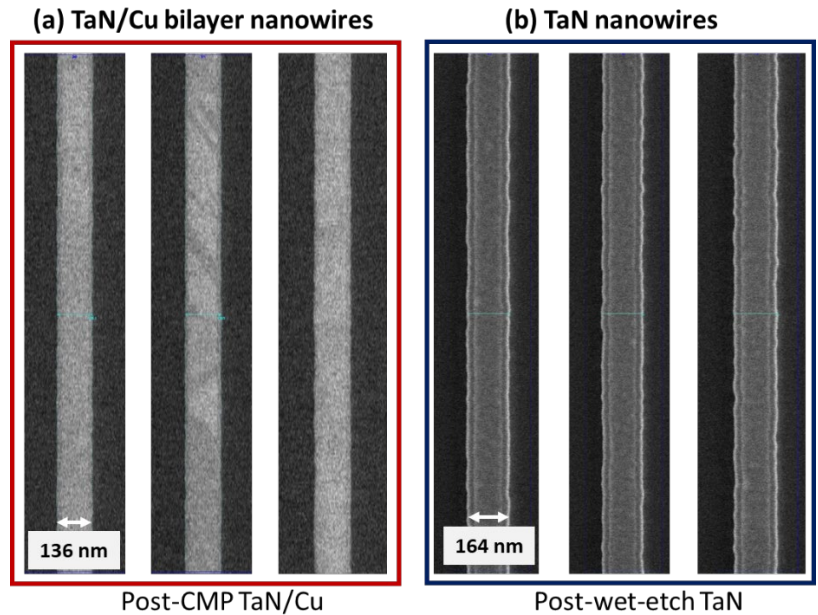

FIG. S1. Representative CD-SEM top-view images of CD test lines measured at three wafer die locations (left to right: (-1, 0), (-4, 6), and (9, 0)). (a) Left (red box): TaN/Cu bilayer CD lines after CMP, showing relatively smooth surfaces with visible Cu grain texture. (b) Right (blue box): TaN CD lines post RIE and hard mask removal (wet etching), exhibiting bright edge contrast consistent with sloped sidewalls from reactive ion etching. Each die measures 12.5 mm×15.5 mm. These sites are three of the 26 CD-SEM sampling locations across the 300 mm wafer (see Fig. 2 in the main text). The scale bar shown in the leftmost image in (a) and (b) applies to all three locations.

## S2. Additional Electrical Measurements

The superconducting transition temperature $T_c$ depends on TaN composition for TaN/Cu bilayer nanowires with a 25-nm TaN layer (Fig. S2). The main manuscript focuses on N/Ta = 0.53. Among the compositions measured, N/Ta = 0.46 (located on the metallic side of the metal-insulator transition) exhibits the highest $T_c$, while the most nitrogen-deficient sample (N/Ta = 0.35) shows no transition above the 1.7 K base temperature of the cryogenic system. Details of the XPS and SIMS characterization, as well as the metal-insulator transition behavior across this composition range are discussed in Ref. [1].

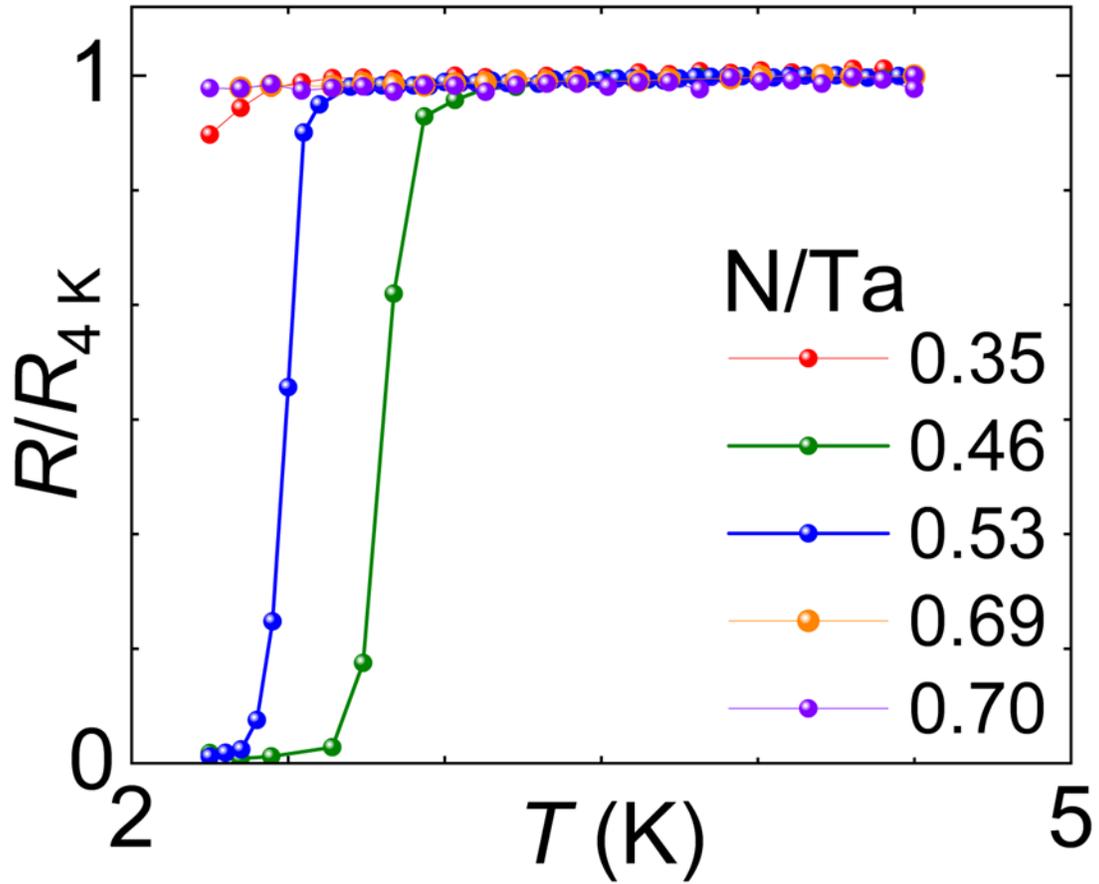

FIG. S2. Normalized resistance $R/R_{4\,K}$ versus temperature for TaN/Cu bilayer nanowires with different N/Ta composition ratios for a TaN thickness of 25 nm and a wire width of 3 μm.

## S3. Determination of $T_c$

The superconducting transition temperature $T_c$ reported in the main text is defined as the midpoint transition temperature, $T_c^{\text{mid}}$, obtained from the peak in $dR/dT$ (Fig. S3(b)), corresponding to the midpoint of the $R(T)$ transition (Fig. S3(a)).

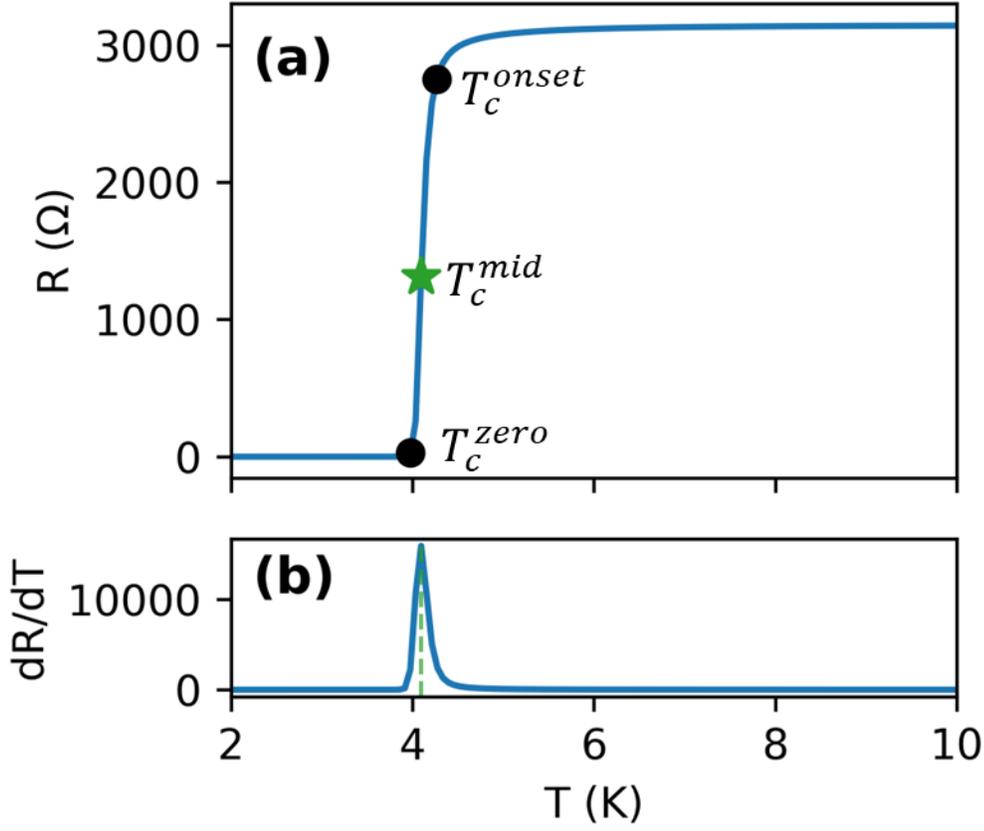

FIG. S3. Extraction of the superconducting transition temperature from $R(T)$ measurements. (a) Resistance as a function of temperature for a representative TaN nanowire with width of 1 μm. Three commonly used definitions of the transition temperature are mentioned: the onset temperature $T_c^{\text{onset}}$ (intersection of the extrapolated normal state resistance with the transition curve), the zero-resistance temperature $T_c^{\text{zero}}$ (where R → 0), and the mid-point temperature $T_c^{\text{mid}}$ (the temperature corresponding to the maximum slope of the transition). (b) Numerical derivative $dR/dT$, where the peak position identifies $T_c^{\text{mid}}$.

References

[1] Bhatia E *et al* 2023 IEEE Trans. Quantum Eng. **4** 5500508